\documentstyle[psfig,twocolumn,aps]{revtex}
\begin{document}

\draft
\title{Character of Atomic Vibrations in a Lennard-Jones Glass}

\author{ Philip B. Allen}
\address{Department of Physics and Astronomy, State University of New York,
Stony Brook, New York 11794-3800}

\author{W. Garber}
\address{Department of Applied Mathematics, State University of New York,
Stony Brook, New York 11794-3600}

\author{L. Angelani}
\address{INFM, SMC-INFM, and Dipartimento di Fisica,
Universit\`a di Roma {\it La Sapienza},
P.le A. Moro 2, 00185 Roma, Italy}

\maketitle

\begin{abstract}

Lennard-Jones glasses (made on a computer by quenching from
liquid state coordinates) are studied in harmonic
approximation.  Vibrational eigenfrequencies and eigenvectors are
found by exact diagonalization for models with periodic boundaries
and N=(500, 2048, and 6980) atoms.  We analyze 
the density of states, mobility edge, and bond-stretching
character of the normal modes.  In agreement with older work
of Grest, Nagel, and Rahman, the upper 7\% of the modes are
localized, and the rest, delocalized.  The modes can not
be differentiated by any property or quantum number except their
eigenfrequency. More specifically, in a given narrow frequency
interval, all modes are globally identical. Transverse or
longitudinal character, for example, disappears.

\end{abstract}


\section{Introduction}

Vibrational properties of disordered media have been reviewed by
various authors \cite{Elliott,Weaire,Visscher,Pohl}.  
Subsequent to most of these reviews, Grest, Nagel, and Rahman (GNR)
\cite{GNR1,GNR2,GNR3,GNR4,GNR5,GNR6} did a careful characterization
of the normal modes of the Lennard-Jones (LJ) glass.
Our previous numerical studies, motivated by heat
conductivity experiments and by theoretical controversy,
focussed on vibrations in amorphous Si (Am-Si)
\cite{Allen}.  Here we present results for a very large LJ
glass model, with local atomic structure very different from
Am-Si.  These results confirm the 
GNR picture.  We use the term
``glass'' simply to mean an amorphous solid, and the
abbreviation am-LJ to denote the LJ glass.

The model studied is a monatomic $6$-$12$ Lennard-Jones 
system, $V(r)=4\epsilon[(\sigma/r)^{12}-(\sigma/r)^6]$,
with potential parameters suitable for Argon:
$\epsilon/k_B = 125.2$ K and $\sigma = 0.3405$ nm.  This potential
gives a satisfactory description of crystalline Ar
\cite{Kittel}.  In its face-centered cubic ({\it
fcc}) structure, each atom is surrounded by 12 nearest neighbors
located a little inside the minimum of the
$V(r)$ at $r=2^{1/6}\sigma$.  The second shell is
more distant by $2^{1/2}$.

Standard molecular dynamics simulations were done
on systems with $N$=$500$, $2048$
and $6980$ atoms, enclosed in a cubic box with periodic boundary
conditions.  A truncated ($r_c = 2.6 \sigma$) LJ potential was used. The
simulated density was $\rho = 42 \times 10^{-3}$ mol/cm$^3$, corresponding
to the reduced value $\rho^*$=$\rho \sigma^3 \simeq 1$.
After an equilibration run at $T = 400$ K using NVE molecular dynamics,
gradual cooling was performed down to a temperature
($T=150$ K) slightly above the melting temperature.
Finally, conjugate gradient minimization found
the glassy minimum configuration.

The resulting structure of am-LJ is
somewhat like that of a smeared-out argon crystal.  The density is
about 5\% smaller. There may be icosahedral aspects in the local
packing \cite{Nelson}, but our models are more similar to the ``LJ
glass'' of Simdyankin {\it et al.} \cite{Simdyankin} than to their
``IC glass'' with strongly icosahedral features.  A shell of
radius r contains on average $(4\pi/3)\rho r^3$ atoms, where
$\rho$ is the number density.  The mean distance from an atom to
its $N$'th neighbor is thus $(3N/4\pi \rho)^{1/3}$.

\par
\begin{figure}[t]
\centerline{\psfig{figure=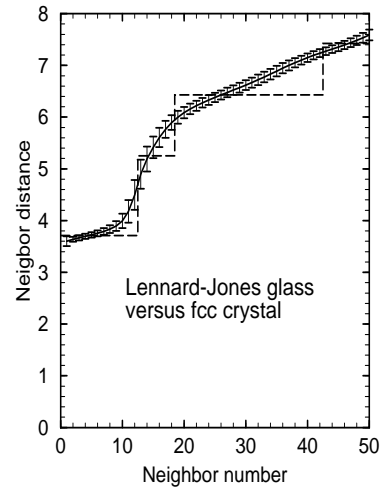,height=2.5in,width=2.5in,angle=0}}
\caption{mean distance to the $N$'th neighbor, for am-LJ (solid
curve) and the corresponding crystal (dashed lines).
Error bars give the variance $<\delta r^2>^{1/2}$ in the
distribution of $N$'th neighbor positions .}
\label{fig:neigh}
\end{figure}
\par

For am-LJ we calculated the mean distance to the $N$'th
neighbor, and the variance $<\delta r^2>^{1/2}$ of this
quantity, which are plotted in Fig. (\ref{fig:neigh}).  The
closest 10 neighbors cluster in a narrow interval near the
distance 3.7\AA \ found in crystalline Ar. The 11th to 13th
neighbors have less well-defined positions. The 15th to 18th atoms
have the least well-defined positions, lying near the second
neighbor shell of the crystal. More distant neighbors start to be
packed as in a random gas, with distance starting to scale like
$N^{1/3}$.

\begin{figure}[t]
\centerline{\psfig{figure=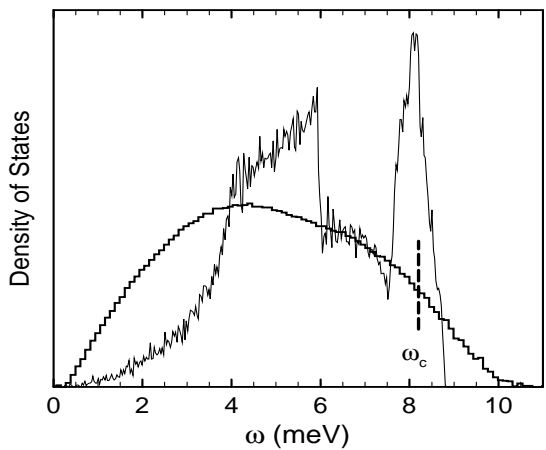,height=2.6in,width=3.4in,angle=0}}
\caption{Density of vibrational eigenstates of am-LJ and the
corresponding {\it fcc} Lennard-Jones (argon-like) crystal.  The
dashed vertical line indicates the position of the mobility edge
in am-LJ.} \label{fig:dos}
\end{figure}

Normal modes of vibration were calculated in harmonic
approximation, using the same periodic boundary conditions
used to compute the positions.  The calculations were done on Njal, 
a Beowulf cluster with 81 dual pentium III processors running at
0.933 GHz. The Scalapack routine PDSYEV was used.  The largest system
size, 6,980 atoms, corresponds to a 
real-symmetric dynamical matrix of dimension 20,940, the largest
that fits conveniently into Njal's memory.  This matrix required about
100 minutes to diagonalize.  The vibrational density of states is
shown in Fig. (\ref{fig:dos}). All three models give essentially
the same result, which is shown here for the 6980-atom model,
and was first found by Rahman et al. \cite{Rahman}.  The
results for the glass are compared with the density
of states of the crystal, which closely resembles
experiment on crystalline Ar \cite{Fujii}. However,
the LJ glass has a surprisingly different density of states, being
featureless, with no 
distinction between longitudinal and transverse modes.

\par
\par

\begin{figure}[t]
\centerline{\psfig{figure=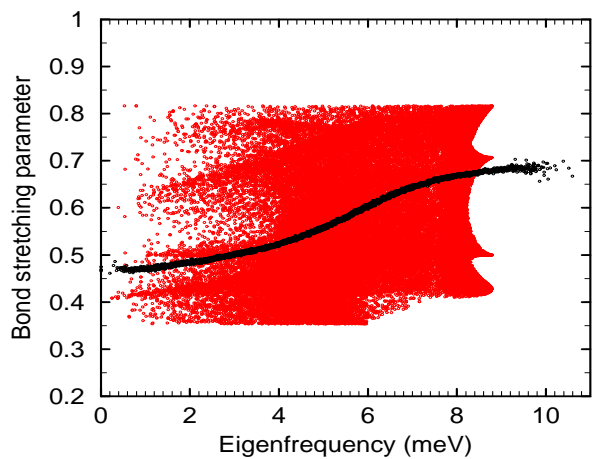,height=2.6in,width=3.4in,angle=0}}
\caption{Bond-stretching character versus frequency for the
normal modes of the glass (narrow dark  band) compared with the
bond-stretching character of the normal modes of the 
crystal (broad light band).} \label{fig:stretch}
\end{figure}

To pursue this issue, it is interesting to look at the
``bond-stretching parameter'' $S_a$ of the $a$'th normal mode,
defined as \cite{Marinov}
\begin{equation}
\\s_a=\left[\frac{\sum_{<i,j>}|(\vec{u}_i^a - \vec{u}_j^a) \cdot
\hat{n}_{ij}|^2}{\sum_{<i,j>}|(\vec{u}_i^a - \vec{u}_j^a)|^2}
\right] ^{\frac{1}{2}} \label{eq:bs}
\end{equation}
where $\vec{u}_i^a$ is the displacement eigenvector on the atom
with position label $i$ and position $\vec{R}_i$.  The unit vector
$\hat{n}_{ij}$ points in the direction $\vec{R}_i -\vec{R}_j$.  In
other words, $S_a$ measures for mode $a$ the relative amount of
stretching of each near-neighbor bond.  For the crystal, this is
rigorously defined by using the 12 bonds from each atom to its
first neighbors.  For the glass, we sum all neighbors whose
separation is less than the mean distance to the 12.5th neighbor.
The results are plotted in Fig. (\ref{fig:stretch}).

\par
\par

This figure shows a very striking difference between the crystal
and the glass.  Crystalline normal modes can be labelled by the
wavevector $\vec{k}$.  In Ar, there are only three normal modes at
each $\vec{k}$.  They have usually quite different bond-stretching
characters, and can normally be classified as longitudinal (L) or
transverse (T).  This leads to a highly differentiated spectrum of
states, seen in the figure as a broad range of values of $S_a$ at
each frequency.  In the glass, the bond-stretching character $S_a$
collapses to a narrow interval whose width diminishes to zero as
$N$ increases to infinity.  Lacking the quantum number $\vec{k}$,
each mode has totally mixed character.  This property was noticed
previously for other glasses \cite{Kamitakahara,Fabian,Allen}.  In
an infinite glass, with infinitely many modes at each frequency,
one could make a unitary rotation (within the subspace of fixed
frequency) to a basis where certain modes would look approximately
pure L in some local region. However, at long enough
distances this character is completely lost.  At low frequencies,
purified modes defined this way could be chosen have a fairly
sharply defined local value of $\vec{k}$. At large enough
distances, the mode would evolve into a spherical mixture of
different directions of $\vec{k}$.  As frequency increases, there
is a crossover (the ``Ioffe-Regel'' (IR) crossover) to the regime
where the local $\vec{k}$ can hardly be defined because it decays
in a distance of one or two neighbors. The IR crossover and other
low-frequency properties will be the subject of a future study.

It was noticed by GNR that a given normal mode, insofar as it
was L, expressed its L character at
specific wavevectors, and insofar as it was T,
expressed that character at other wavevectors.  This enabled
them to draw pseudo-dispersion curves showing very different
character for L and T aspects.  This
does not contradict our view that a given mode cannot be
classified.  Each mode is simultaneously L and T, probably even
at the same point in space unless purified as described above.

\par
\begin{figure}[t]
\centerline{\psfig{figure=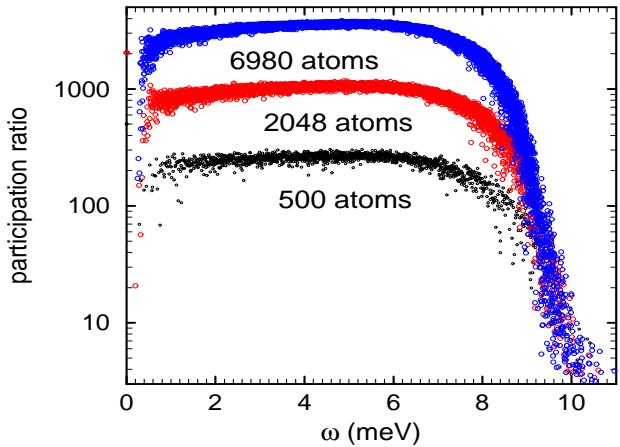,height=2.6in,width=3.4in,angle=0}}
\caption{Inverse participation ratio on logarithmic scale versus
eigenfrequency for all normal modes of systems of various sizes.}
\label{fig:iprlog}
\end{figure}
\par

In principle, normal modes of vibration, just like electron
eigenstates in quantum theory, can be rigorously classified as
extended (E) or localized (L).  In three dimensions the
vibrational spectrum has sharp E/L boundary frequencies
(``mobility edges,'' $\omega_c$) separating these two kinds of
modes.  The IR crossover is low in the
spectrum of a glass, but the E/L boundary is usually high in the
spectrum. 

An hypothesis that the mobility edge coincides with the IR
crossover \cite{Alexander} has not proved true for any model that
has been studied numerically, but has remained
a popular misconception.  Our large system provides a useful
confirmation that previous studies have correctly located the
mobility edge.  We compute the
participation ratio $p_a$ for each normal mode, defined as
\begin{equation}
p_a = \frac{[\sum_i |\vec{u}_i^a|^2]^2}{\sum_i |\vec{u}_i^a|^4}.
\label{eq:pr}
\end{equation}
The values of $p_a$ are plotted on a logarithmic scale versus
eigenfrequency for all three models in Fig. (\ref{fig:iprlog}).

Notice first that the value of $p_a$ hovers at slightly higher
than 0.5$N$ for a significant part of the spectrum (frequencies
2.5 to 6.5 meV).  These are very well delocalized states whose
amplitude does not vary too much from atom to atom.  At low
frequencies (below 1 meV for the 6980-atom model, and pushing up
to 2 meV for the 500-atom model) there is a trend to lower values
of $p_a$, and there occur some idiosyncratic modes with $p_a$
reduced by 10 or more.  Their interpretation remains difficult
\cite{Feldman}.  At least some of this behavior has to be
attributed to finite-size artifacts.  We hope to make a detailed
study later. Finally, at high
frequencies, very low values of $p_a$ occur.  These states are
strongly Anderson localized.  The mobility edge is estimated to be
at $\omega_c=8.2 \pm 0.1$ meV.  This estimate was found by fitting
a straight line to the logarithmic data at the point where $p_a
\approx 0.25N$.  Extrapolating this line back to the horizontal
line $p_a=0.5N$, the intersection occurs at around 8.2 meV, with
uncertainty which diminishes as the system size increases.  This
value of $\omega_c$ corresponds to having the lower 93\% of the
spectrum delocalized, and the upper 7\% localized, and agrees
with the result of GNR \cite{GNR1,GNR6}.

These results make an interesting comparison to our earlier work
on amorphous Si \cite{Allen}.  One major difference is that the
density of vibrational states in am-Si retains much of the structure seen
in the crystal, with acoustic and optical types of vibrations
separated by a region of low density of states.  To the contrary,
in am-LJ, the density of states loses differentiation and
is completely featureless.  Apart from this, other
properties of am-LJ vibrations are reminiscent of those in
am-Si.  We see similar behavior in the collapse of the
bond-stretching character onto a single frequency-dependent curve.
The anomalous states at very low frequency, and the strongly
localized states at very high frequency, have similar character.
In both glasses, fewer than 10\% of the modes are localized,
always at the upper end of the spectrum.

\acknowledgements

We thank G. S. Grest, J. Fabian, J. L. Feldman and G. Viliani for help.
Work at Stony Brook was supported by NSF grant DMR-0089492.  We
thank K. K. Likharev for the use of the
Njal computer cluster, which was purchased using a grant from DOD's DURIP
program.


\begin{references}

\bibitem{Elliott}       R. J. Elliott and P. L. Leath,
                        in {\sl Dynamical Properties of Solids},
                        (eds.  G. K. Horton and A. A. Maradudin),
                        North Holland, Amsterdam, 1975; vol. 2, p.385.

\bibitem{Weaire}        D. Weaire and P. C. Taylor,
                        in {\sl Dynamical Properties of Solids},
                        (eds.  G. K. Horton and A. A. Maradudin),
                        North Holland, Amsterdam, 1980; vol. 4, p.1.

\bibitem{Visscher}      W. M. Visscher and J. E. Gubernatis,
                        in {\sl Dynamical Properties of Solids},
                        (eds.  G. K. Horton and A. A. Maradudin),
                        North Holland, Amsterdam, 1980; vol. 4, p.63.

\bibitem{Pohl}          R. O. Pohl,
                        in {\sl Encylopedia of Applied Physics},
                        Wiley-VCH, 1998, vol. 23, p.223.

\bibitem{GNR1}		S. R. Nagel, A. Rahman, and G. S. Grest,
			Phys. Rev. Lett. {\bf 47}, 1665 (1981).

\bibitem{GNR2}		G. S. Grest, S. R. Nagel, and A. Rahman,
			Phys. Rev. Lett. {\bf 49}, 1271 (1982).

\bibitem{GNR3}		G. S. Grest, S. R. Nagel, and A. Rahman,
                        Phys. Rev. B {\bf 29}, 5968 (1984).

\bibitem{GNR4}		S. R. Nagel, G. S. Grest, and A. Rahman,
			Phys. Rev. Lett. {\bf 53}, 368 (1984).

\bibitem{GNR5}		S. R. Nagel, G. S. Grest, S. Feng, and L. M. Schwartz,
			Phys. Rev. B {\bf 34}, 8667 (1986).

\bibitem{GNR6}		S. R. Nagel, G. S. Grest, and A. Rahman,
                        Physics Today, Oct. 1983, p. 24.

\bibitem{Allen}         P. B. Allen, J. L. Feldman, J. Fabian, and F. Wooten,
                        Phil. Mag. B {\bf 79}, 1715-32 (1999).

\bibitem{Kittel}        C. Kittel,
                        {\it Introduction to Solid State Physics},
                        7th Edition, Wiley, New York, 1996.

\bibitem{Nelson}        D. R. Nelson and F. Spaepen,
                        Solid State Phys. {\bf 42}, 1 (1989).

\bibitem{Simdyankin}    S. I. Simdyankin, M. Dzugutov,
                        S. N. Taraskin, and S. R. Elliott,
                        Phys. Rev. B {\bf 63}, 184301 (2001).

\bibitem{Rahman}	A. Rahman, M. J. Mandell, and J. P. McTague,
			J. Chem. Phys. {\bf 64}, 1564 (1976).

\bibitem{Fujii}          Y. Fujii, N. A. Lurie, R. Pynn, and G. Shirane,
                        Phys. Rev. B {\bf 10}, 3647 (1974).

\bibitem{Marinov}       M. Marinov and N. Zotov,
                        Phys. Rev. B {\bf 55}, 2938 (1997).

\bibitem{Kamitakahara}  W. A. Kamitakahara, C. M. Soukoulis, H. R. Shanks,
                        U. Buchenau, and G. S. Grest,
                        Phys. Rev. B {\bf 36}, 6539 (1987).

\bibitem{Fabian}        J. Fabian and P. B. Allen,
                        Phys. Rev. Letters {\bf 79}, 1885 (1997).

\bibitem{Alexander}     S. Alexander,
                        Phys. Rev. B {\bf 40}, 7953 (1989).

\bibitem{Feldman}       J. L. Feldman, P. B. Allen, and S. R. Bickham,
                        Phys. Rev. B {\bf 59}, 3551 (1999).

\end{references}
\end{document}